\documentclass[twocolumn,aps,preprintnumbers,amsmath,amssymb,superscriptaddress,nofootinbib,longbibliography]{revtex4}

\usepackage{graphicx,subfigure}
\usepackage{longtable}
\usepackage{dcolumn}
\usepackage{bm}
\usepackage[all]{xy}
\usepackage{color}
 \usepackage{graphicx}
\usepackage[usenames,dvipsnames]{xcolor}
\usepackage[unicode=true,pdfusetitle,
 bookmarks=true,bookmarksnumbered=false,bookmarksopen=false,citecolor=Turquoise,
 breaklinks=false,pdfborder={0 0 1},backref=false,colorlinks=true,pdfpagemode=FullScreen]
 {hyperref}

\usepackage{lscape}
\usepackage{tikz}
\usepackage{pgf}
\usepackage{color}
\usepackage{booktabs}
\usepackage{multirow}
\usepackage{siunitx}

\usepackage{lipsum}

\makeatletter

\newcommand{\fmslash}[2][0mu]{%
  \mathchoice
    {\fmsl@sh\displaystyle{#1}{#2}}%
    {\fmsl@sh\textstyle{#1}{#2}}%
    {\fmsl@sh\scriptstyle{#1}{#2}}%
    {\fmsl@sh\scriptscriptstyle{#1}{#2}}}
\newcommand{\fmsl@sh}[3]{%
  \m@th\ooalign{$\hfil#1\mkern#2/\hfil$\crcr$#1#3$}}
\makeatother

\newcommand{\beq}{\begin{equation}}
\newcommand{\eeq}{\end{equation}}
\newcommand{\bea}{\begin{eqnarray}}
\newcommand{\eea}{\end{eqnarray}}
\mathchardef\minus="002D

\begin{document}
\title{Hadronic Top Quark Polarimetry with ParticleNet}

\author{Zhongtian Dong}
\email{cdong@ku.edu}
\affiliation{Department of Physics and Astronomy, University of Kansas, Lawrence, KS 66045, USA}

\author{Dorival Gon\c{c}alves}
\email{dorival@okstate.edu}
\affiliation{Department of Physics, Oklahoma State University, Stillwater, OK, 74078, USA}

\author{Kyoungchul Kong}
\email{kckong@ku.edu}
\affiliation{Department of Physics and Astronomy, University of Kansas, Lawrence, KS 66045, USA}

\author{Andrew J.~Larkoski}
\email{larkoa@gmail.com}
\affiliation{Department of Physics and Astronomy, University of California, Los Angeles, CA 90095, USA \\
Mani L. Bhaumik Institute for Theoretical Physics, University of California, Los Angeles, CA 90095, USA}

\author{Alberto Navarro}
\email{alberto.navarro\_serratos@okstate.edu}
\affiliation{Department of Physics, Oklahoma State University, Stillwater, OK, 74078, USA}

\begin{abstract}
\noindent 
Precision studies for top quark physics are a cornerstone of the Large Hadron Collider  program. Polarization, probed through decay kinematics, provides a unique tool to scrutinize the top quark across its various production modes and to explore potential new physics effects. However, the top quark most often decays hadronically, for which unambiguous identification of its decay products sensitive to top quark polarization is not possible. In this Letter, we introduce a jet flavor tagging method to significantly improve spin analyzing power in hadronic decays, going beyond exclusive kinematic information employed in previous studies.  We provide parametric estimates of the improvement from flavor tagging with any set of measured observables and demonstrate this in practice on simulated data using a Graph Neural Network (GNN). We find that the spin analyzing power in hadronic decays can improve by approximately 20\% (40\%) compared to the kinematic approach, assuming an efficiency of 0.5 (0.2) for the network.
\end{abstract}


\maketitle

\medskip
\noindent {\bf I.~Introduction} 

\noindent As the most massive fundamental particle, the top quark is especially sensitive to physics at the weak scale and beyond through its large Yukawa coupling~\cite{Carena:1993bs,PhysRevD.75.015002,PhysRevD.80.076002,Panico:2011pw,Matsedonskyi:2012ym,Pomarol:2012qf,PhysRevD.84.015003,Bellazzini:2014yua,Panico:2015jxa}. Furthermore, its mass plays a crucial role in the stability of the universe~\cite{Degrassi:2012ry,Bezrukov:2012sa}. When studying top quark physics, polarization is a valuable tool. Since the top quark decays before it hadronizes and before depolarization by soft QCD takes place, its polarization can be probed through the kinematics of its decays. This unique perspective on top quark production, viewing it not just as a raw rate but as a collection of distinct polarized processes, has been repeatedly leveraged in new physics searches across the leading production channels for top quark at the LHC ($t\bar{t}$, single-top, $t\bar{t}W$, and $t\bar{t}Z$)~\cite{Bernreuther:1993hq,Arai:2007ts,Perelstein:2008zt,Krohn:2011tw,Barger:2011pu,Baumgart:2011wk,Buckley:2015vsa,Buckley:2015ctj,Goncalves:2016qhh,Goncalves:2018agy,Barger:2019ccj,Goncalves:2021dcu,Barman:2021yfh,MammenAbraham:2022yxp,Maltoni:2024tul}, studies of spin correlations in $t\bar{t}$ production~\cite{Mahlon:1995zn,Mahlon:2010gw,CDF:2010yag,D0:2011psp,ATLAS:2019zrq,CMS:2019nrx,Aguilar-Saavedra:2021ngj,Aguilar-Saavedra:2022kgy}, and more recently in the observation of entanglement in $t\bar{t}$ production~\cite{Afik:2020onf,Fabbrichesi:2021npl,Severi:2021cnj,Aguilar-Saavedra:2022uye,Afik:2022kwm,Afik:2022dgh,Severi:2022qjy,Aoude:2022imd,PhysRevD.109.115023,Cheng:2023qmz,ATLAS:2023fsd,CMS-PAS-TOP-23-001,CMS:2024vqh}.

However, most of these studies have focused on its leptonic decays, where the charged lepton from the subsequent $W$ boson decay is perfectly correlated with the spin of the top quark \cite{Bernreuther:2010ny}. Experimentally observing electrons or muons is very clean, allowing for precise measurements.  However, top quark pairs only decay to electrons and muons about 5\% of the time, leading to a significant statistical limitation. In contrast, top quark pairs decay hadronically or semi-leptonically  nearly 95\% of the time. In particular, this allows for further exploration of the phenomenologically motivated highly boosted regime~\cite{PhysRevD.109.115023}, but it comes at the cost of a more challenging final state to analyze.  While, in a hadronic decay, the down-type quark from subsequent $W$ decay is perfectly correlated with the spin of the top quark, subsequent parton showering and hadronization wash away any unambiguous identification of the down quark.\footnote{While top quarks are not polarized in $t\bar{t}$ production at the LHC due to its QCD dominated nature, they display spin correlations. In spin correlation studies, the down-type jet can serve as a proxy for the top quark's spin.} 

In this Letter, we explore jet flavor discrimination observables to improve sensitivity to top quark polarization in its hadronic decays.  This problem has a long history~\cite{Jezabek:1994qs}, with significant improvement in spin resolving power identified in Ref.~\cite{Tweedie:2014yda}.  Due to the left-handed nature of the weak decay of the top quark, the probabilities that the down-type quark from $W$ decay is the harder or softer jet are not equal. Thus, using a probability-weighted vector enhances sensitivity to top quark polarization.  This optimal direction, which relies solely on kinematic information about the jets, is given by  
\begin{align}
\label{eq:opt-kin}    
\hspace{-0.12cm} {\vec q}_\text{opt}^\text{\hspace{-0.01cm} kin} =  p(d\to q_\text{hard}|c_W)\hat q_\text{hard}
+p(d\to q_\text{soft}|c_W)\hat q_\text{soft}\,,                      
\end{align}
where $c_W$ is the cosine of the helicity angle, the angle between the down-type quark and the opposite direction of the bottom quark $(\hat{z}=-\hat{b})$ from the top decay, in the frame of the $W$ boson. $\hat q_\text{hard}$ ($\hat q_\text{soft}$) is the unit vector in the direction of the harder (softer) of the two jets from $W$ boson decay in the rest frame of the top quark.  The length of this vector is a measure of the spin analyzing power, which was evaluated at leading order to be $\beta_\text{opt}^\text{\hspace{-0.03cm} kin} \equiv \langle |\vec q_\text{opt}^\text{ kin}|\rangle\approx 0.64$, averaged over the helicity angle.

Of course, there are many more qualities of up-type and down-type jets that differ than just their kinematic distributions, such as 
their jet charges~\cite{Field:1977fa,Krohn:2012fg,Fraser:2018ieu,Kang:2023ptt}, or their hadronic particle content. 
Ignoring this information potentially misses out on significant improvements to the spin analyzing power.  There has been speculation in the literature that the improvement from using additional information would be mild~\cite{Tweedie:2014yda,Han:2023fci}, but no study of its effect has ever been completed to validate or refute this assumption.  Here, we address this question directly, including myriad information beyond just jet kinematics for flavor tagging, establishing a spin analyzing power substantially better than that from Ref.~\cite{Tweedie:2014yda}.  While we will merely assume that jet flavor is well-defined and corresponds to that of the initiating particle from a leading-order parton shower simulation, there have been significant developments in establishing robust and theoretically well-defined jet flavor definitions recently~\cite{Banfi:2006hf,Caletti:2022hnc,Caletti:2022glq,Czakon:2022wam,Gauld:2022lem,Caola:2023wpj}.

\medskip
\noindent {\bf II.~Theoretical framework} 

\noindent 
Because of its short lifetime, the top quark decays before it hadronizes and spin decorrelations effects occur~\cite{Mahlon:2010gw,ParticleDataGroup:2022pth}. This ensures that the final states of the top quark are correlated with its polarization axis as
\begin{align}
    \frac{1}{\Gamma}\frac{d\Gamma}{d\cos\theta_i}=\frac{1}{2}\left( 1+ \beta_i p\cos\theta_i \right)\,,
    \label{eq:beta}
\end{align}
where $\Gamma$ represents the partial decay width, $\theta_i$ is the angle between the final state particle $i$ and top quark spin axis in the top quark rest frame, $p$ denotes the degree of polarization of the ensemble, and $\beta_i$ is the spin analyzing power for the decay product $i$~\cite{Bernreuther:2010ny}. The spin analyzing power is maximal for charged leptons and down quarks, $\beta_{\ell^+,\bar{d}}=1$. However, tagging a $d$-quark in a collider environment is challenging. One possible solution is to use the softest of the two light jets from the top decay in the top quark rest frame, which yields $\beta_{\rm soft} \simeq 0.5$~\cite{Jezabek:1994qs}. This can be improved by using the direction defined in Eq.~(\ref{eq:opt-kin}) as a proxy for the $d$-quark, leading to $\beta_\text{opt}^\text{\hspace{-0.03cm} kin}\simeq 0.64$~\cite{Tweedie:2014yda}.

To improve the hadronic top polarimetry, we first establish a parametric estimate for the increase of the spin analyzing power due to the measurement of a collection of non-kinematic observables $\{{\cal O}\}$ -- such as jet charge, particle multiplicity, or the like -- in addition to the helicity angle.  Without oracle knowledge of which subjet corresponds to the down quark, the direction most sensitive to the polarization of the top quark is now
\begin{align}
\label{eq:opt}    
\vec q_\text{opt} =& \,\, p(d\to q_\text{hard}|c_W,\{{\cal O}\}) \, \hat q_\text{hard}\\
&+p(d\to q_\text{soft}|c_W,\{{\cal O}\}) \, \hat q_\text{soft}\,.\nonumber
\end{align}
We will assume that these non-kinematic observables are independent of the kinematics of the jets, which is a good approximation because the jets are produced from on-shell decay of the $W$ boson in the narrow-width approximation and so their dynamics are Lorentz-invariant.  For simplicity of resulting expressions, we will work with the squared-length of this vector as a measure of the sensitivity to the top quark's polarization, where
\begin{align}
&|\vec q_\text{opt}|^2 = 1 - 2p(d\to q_\text{hard}|c_W,\{{\cal O}\})p(d\to q_\text{soft}|c_W,\{{\cal O}\})\nonumber\\
&\hspace{4cm}\times\left(
1-\hat q_\text{hard}\cdot \hat q_\text{soft}
\right)\,.
\end{align}
The dot product of the two unit vectors, $\hat q_\text{hard}\cdot \hat q_\text{soft}$, is exclusively a function of the helicity angle, because in the top quark rest frame, the $W$ boson has a fixed energy.  However, the prefactor product of probabilities will be modified from its exclusively kinematic form identified in Ref.~\cite{Tweedie:2014yda}.

Using the definition of conditional probability, we have
\begin{align}
p(d\to q_\text{hard}|c_W,\{{\cal O}\}) = \frac{p(d\to q_\text{hard},\{{\cal O}\}|c_W)}{p(\{{\cal O}\}|c_W)}\,,
\end{align}
where the denominator can be expressed as
\begin{align}
p(\{{\cal O}\}|c_W) &= p(\{{\cal O}\}|d\to q_\text{hard})p(d\to q_\text{hard}|c_W)\\
&\hspace{1cm}+p(\{{\cal O}\}|d\to q_\text{soft})p(d\to q_\text{soft}|c_W)\,.\nonumber
\end{align}
Using the assumption that the observables $\{{\cal O}\}$ are independent of the helicity angle, the numerator can be expressed as
\begin{align}
p(d\to q_\text{hard},\{{\cal O}\}|c_W) 
= p(\{{\cal O}\}|d\to q_\text{hard})p(d\to q_\text{hard}|c_W).
\end{align}
With these results, the squared length of the optimal polarization vector can be written as
\begin{align}
&|\vec q_\text{opt}|^2 = 1  
- \frac{2{\cal L}_{\cal O}\, p(d\to q_\text{hard}|c_W)p(d\to q_\text{soft}|c_W)}{\left(p(d\to q_\text{hard}|c_W)+{\cal L}_{\cal O}p(d\to q_\text{soft}|c_W)\right)^2} \nonumber \\
& \hspace{2cm }\times \left(
1-\hat q_\text{hard}\cdot \hat q_\text{soft}
\right).
\end{align}
Here, ${\cal L}_{\cal O}$ is the likelihood ratio of the non-kinematic observables:
\begin{align}
{\cal L}_{\cal O}\equiv \frac{p(\{{\cal O}\}|d\to q_\text{soft})}{p(\{{\cal O}\}|d\to q_\text{hard})}\,.
\end{align}

Now, we would like to compare this squared length with the case where no additional observables $\{{\cal O}\}$ are measured.  To do this, we will integrate over the non-kinematic observables, defining the average squared length still dependent on the helicity angle, as
\begin{align}
\langle |\vec q_\text{opt}|^2\rangle_{\cal O} &= \int d{\cal L}_{\cal O}\, p({\cal L}_{\cal O})\, |\vec q_\text{opt}|^2\,.
\end{align}
Note that the probability distribution of the likelihood is
\begin{align}
&p({\cal L}_{\cal O}) = p({\cal L}_{\cal O}|d\to q_\text{hard})\\
&\hspace{1.5cm}\times\left[
p(d\to q_\text{hard}|c_W)+{\cal L}_{\cal O}p(d\to q_\text{soft}|c_W)
\right]\,,\nonumber
\end{align}
using the definition of the likelihood.  The squared optimal vector length, averaged over observables, is then
\begin{widetext}
\begin{align}\label{eq:obsave}
\langle |\vec q_\text{opt}|^2\rangle_{\cal O} 
&=1-2p(d\to q_\text{hard}|c_W)p(d\to q_\text{soft}|c_W)\left(
1-\hat q_\text{hard}\cdot \hat q_\text{soft}
\right)\\
&\hspace{1cm}\times\left[
p(d\to q_\text{hard}|c_W)\int d{\cal L}_{\cal O}\, \frac{{\cal L}_{\cal O}\,p({\cal L}_{\cal O}|d\to q_\text{hard})}{\left(p(d\to q_\text{hard}|c_W)+{\cal L}_{\cal O}p(d\to q_\text{soft}|c_W)\right)^2}\right.\nonumber\\
&\hspace{2cm}\left.+\,p(d\to q_\text{soft}|c_W)\int d{\cal L}_{\cal O}\, \frac{{\cal L}_{\cal O}^2\,p({\cal L}_{\cal O}|d\to q_\text{hard})}{\left(p(d\to q_\text{hard}|c_W)+{\cal L}_{\cal O}p(d\to q_\text{soft}|c_W)\right)^2}
\right]\nonumber\,.
\end{align}
\end{widetext}

Progress can be made on actually evaluating this integral with no more assumptions on the structure of the probability $p({\cal L}_{\cal O})$.  First, we note that because $p({\cal L}_{\cal O})$ and $p({\cal L}_{\cal O}|d\to q_\text{hard})$ are both normalized, the mean value of $p({\cal L}_{\cal O}|d\to q_\text{hard})$ is 1:
\begin{align}
\int d{\cal L}_{\cal O}\, p({\cal L}_{\cal O}|d\to q_\text{hard})\, {\cal L}_{\cal O} = 1\,.
\end{align}
Then, we can express the probability in a moment expansion as (see, e.g., Ref.~\cite{Kang:2023ptt}) 
\begin{align}
p({\cal L}_{\cal O}|d\to q_\text{hard}) &= \delta({\cal L}_{\cal O}-1) + \frac{\sigma^2}{2}\delta''({\cal L}_{\cal O}-1)+\cdots\,,
\end{align}
where higher central moments are implicit in the ellipses.  Here, $\sigma^2$ is the variance 
of $p({\cal L}_{\cal O}|d\to q_\text{hard})$ which is necessarily non-negative. 
Using this expansion, we can then perform the integral in Eq.~\eqref{eq:obsave}:
\begin{align}
&\langle |\vec q_\text{opt}|^2\rangle_{\cal O} =1\\
&-2\,p(d\to q_\text{hard}|c_W)p(d\to q_\text{soft}|c_W)\left(
1-\hat q_\text{hard}\cdot \hat q_\text{soft}
\right)\nonumber\\
&+2\sigma^2 \,p(d\to q_\text{hard}|c_W)^2p(d\to q_\text{soft}|c_W)^2\left(
1-\hat q_\text{hard}\cdot \hat q_\text{soft}
\right)\nonumber\\
&+\cdots\nonumber\,.
\end{align}
Note that the first and second lines on the right-hand side of this expression are the squared optimal vector length only measuring the helicity angle $c_W$.  The third 
line includes the effect of making additional non-kinematic measurements, and is explicitly positive, demonstrating that additional measurements necessarily improve the polarimeter.

To concretely determine the size of improvement, we can further integrate over helicity angle $c_W$, given the expressions for the probabilities from leading-order top quark decay~\cite{Tweedie:2014yda,PhysRevD.109.115023}.  
With the PDG values for the top, bottom,  and $W$ masses~\cite{ParticleDataGroup:2022pth}, we find  
the following expansion as a measure of the spin analyzing power\footnote{We note that  $\sqrt{\langle |\vec q_\text{opt}^\text{\hspace{0.007cm} kin}|^2\rangle}\approx 0.643$ is
ever so slightly larger than the mean spin resolving power, $\langle |\vec q_\text{opt}^\text{\hspace{0.01cm} kin}|\rangle\approx 0.640$.}
\begin{align}
\label{eq:sigma}
\sqrt{\langle |\vec q_\text{opt}|^2\rangle} &\approx 0.643 + 0.163\sigma^2 
+\cdots\,.
\end{align}
That is, the top quark spin resolving power can be improved beyond the result of Ref.~\cite{Tweedie:2014yda} by at least 10\% of the variance of the likelihood distribution from the measurement of non-kinematic observables.  In general, we expect that the distribution $p({\cal L}_{\cal O}|d\to q_\text{hard})$ is peaked around ${\cal L}_{\cal O} = 0$, with a tail extending to large values of ${\cal L}_{\cal O}$, such that the mean is 1.  As such, the variance of this distribution can be relatively large compared to the mean, $\sigma^2\sim 1$, so we expect that by making additional measurements, the spin analyzing power in hadronic top decays can approach or even exceed 0.8. 

Motivated by the potential improvements shown in this analytical derivation,  we will demonstrate how a realistic analysis can incorporate attainable observables to probe the hadronic top quark polarization. To achieve this, we will leverage the capabilities of machine learning techniques.

\medskip
\noindent {\bf III.~Analysis} 

\noindent 
In this analysis, we aim to access the improvements on hadronic top quark polarimetry by going beyond the procedure presented in Eq.~\eqref{eq:opt-kin}, which accounts for the kinematics for the two light subjets from the top quark decay. To achieve this, we will perform two additional calculations. First, we will compute the spin analyzing power using the kinematic features of the three subjets from the hadronic top quark, using a Deep Neural Network (DNN). Since the leading kinematic features are encoded in Eq.~\eqref{eq:opt-kin}, we expect this approach to yield results similar to those obtained from this equation.
We will contrast this calculation with an analysis that further explores the jet substructure, including additional information from the constituents of the subjets, such as kinematics, charges, and particle identification (PID) when realistically attainable, using a Graph Neural Network (GNN). 

We started the analysis by generating three samples for semi-leptonic top pair production, $pp\to t \bar{t}\to \ell^\pm \nu 2b 2j$ within the Standard Model
at the $\sqrt{s}=14$~TeV LHC, where $\ell^\pm=e^\pm$ or $\mu^\pm$, using \texttt{MadGraph5\_aMC@NLO}~\cite{Alwall:2014hca,BuarqueFranzosi:2019boy}. We use \texttt{NNPDF2.3QED} for parton distribution function~\cite{Ball:2013hta}, setting the factorization and renormalization scales to $\mu_F=\mu_R=\left(\sqrt{m_t^2+p_{ T t}^2}+\sqrt{m_t^2+p_{T \bar{t}}^2} \right)/2$.  The first two samples were generated with  left and right-handed polarized hadronic top quarks. The third sample was generated with unpolarized top quarks. No kinematic selections on the final state particles were applied at the generation level, except for $p_{Tt}>200$ GeV. The events were then passed through \texttt{PYTHIA8} to simulate parton shower and hadronization effects~\cite{Bierlich:2022pfr}. 
		    
For event reconstruction, we begin by requiring all final state particles to satisfy $|\eta|<3$ and $p_T>1$~GeV. We then cluster the hadronic activity with the Cambridge/Aachen (CA) algorithm with $R=1.5$~\cite{Cacciari:2011ma}. We require at least one fatjet with $p_{TJ}>250$~GeV.  
For the jet substructure analysis~\cite{Butterworth:2008iy,Plehn:2010st}, we begin by declustering the fatjet until we obtain at least three subjets with $m_\text{subjet}<30$~GeV. If there are more than three subjets, we consider the hardest four. If a fourth jet is present but is softer than the third one by a factor of three or more, we disregard it. We then  reconstruct the hadronic top quark by combining the two hardest jets with the third and fourth, if present, keeping the combination that yields a mass closest to the top quark mass, $m_{\text{fatjet}}=[165,190]$~GeV. For each subjet, we record the $\left(p_T,\,\eta,\,\phi,\,E,\,\text{charge},\,\text{PID}\right)$ of up to 40 constituents (shown in the bottom panel of Table \ref{table:variables}). 

All events are then processed through a ``matching'' procedure, in which each jet is assigned to a parton. We compute the $\Delta R$ distance between each jet and parton, assigning each jet to the closest parton. If a parton cannot be uniquely matched to a single jet, the event is discarded. 

\textbf{\begin{table}[!t]
\centering
\renewcommand\arraystretch{1.1}
\begin{tabular}{ c | l } 
 \hline
Variable & \hspace{2.2cm}Definition\\ 
 \hline\hline
$\Delta \eta_{t}$ & ~~difference in pseudorapidity between \\ & ~~the particle and the top jet axis\\ 
$\Delta \phi_{t}$ & ~~difference in azimuthal angle between\\ & ~~the particle and the top jet axis\\ 
 \hline
$\Delta \eta_{j}$ & ~~difference in pseudorapidity between \\ & ~~the particle and the subjet axis\\ 
$\Delta \phi_{j}$ & ~~difference in azimuthal angle between \\ & ~~the particle and the subjet axis\\ 
log $p_T$ & ~~logarithm of the particle’s $p_T$\\ 
log $E$ & ~~logarithm of the particle’s Energy\\ 
$q$ & ~~electric charge of the particle\\  
isElectron & ~~if the particle is an electron\\ 
isMuon & ~~if the particle is a muon\\ 
isPhoton & ~~if the particle is a photon\\ 
isChargedHadron & ~~if the particle is a charged hadron\\ 
isNeutralHadron & ~~if the particle is a neutral hadron\\ 
 \hline
\end{tabular}
\caption{Input features in GNN: the positions of points in the graph (top) and the attributes of each particle (bottom).
}
\label{table:variables}
\end{table}}

For the DNN implementation, we input the four-momenta as ($p_T, \eta, \phi, E$) of the $b$-jet, harder jet, and softer jet in that order. Additionally, the helicity angle in the top rest frame is included as an input feature, resulting in a total of 13 input features. The binary label given to each event indicates whether the harder jet corresponds to the down-type jet. The network architecture consists of three hidden layers, each with $32$ dimensions with RELU activation function. The output layer is one-dimensional with a sigmoid activation function. 

\begin{figure}[bt!]
    \centering
    \includegraphics[width=1\linewidth,clip]{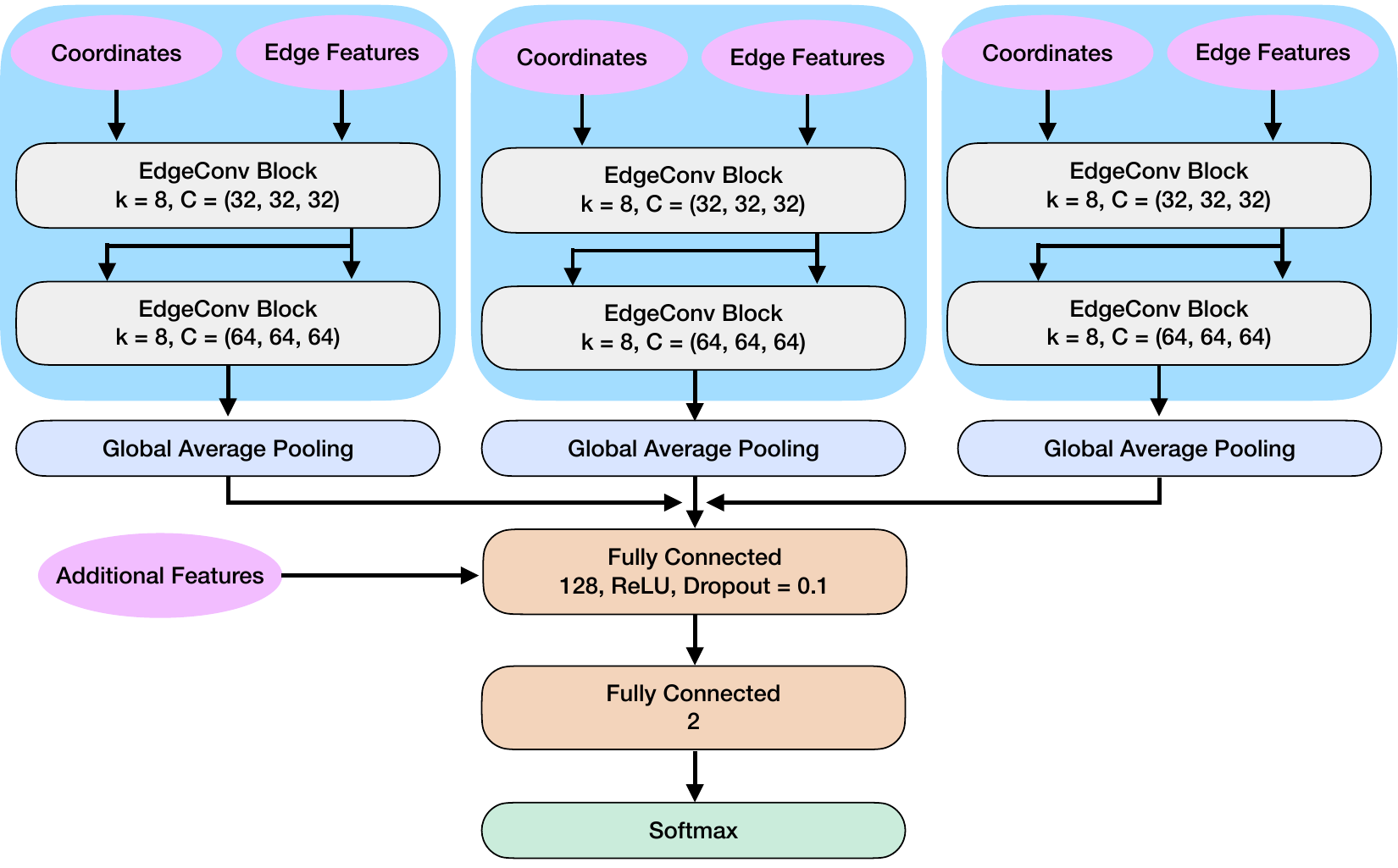}
    \caption{The architecture of the modified version of ParticleNet \cite{Qu:2019gqs}. Three separate graphs, each representing a subjet, are fed into three distinct blocks in the network. After processing, the graphs are pooled, and concatenated, then passed through linear layers before producing the final output. In the diagram, $k$ denotes the number of nearest neighbors, and $C$ represents the number of channels. 
\label{fig:particlenetlight}}
\end{figure}

For the GNN implementation, we modified the ParticleNet architecture described in Ref.~\cite{Qu:2019gqs}, maintaining the same Edge Convolution blocks. Unlike the original design, which processes constituents from one jet at a time to produce a label, our modified GNN takes inputs from the constituents of three jets as three separate graphs. It performs three distinct graph convolutions and combines the resulting information to produce a single label. This modification is essential for our goal of identifying the down-flavor subjet within the hadronic top quark fatjet. The coordinates of the subjet constituents in the graph are then defined as the $\eta$ and $\phi$ relative to the top fatjet axis,  and we include up to $40$ constituents for each subjet. The input features are summarized in Table~\ref{table:variables}.
After the Edge Convolution blocks, each graph is pooled and flattened in the same manner, then concatenated into a single linear input of total dimension of 192. The helicity angle is included as a supplementary feature, by concatenating it with the linear layer following the edge convolutions. 
The combined inputs are then fed into a fully connected linear layer with 128 neurons before the output layer. The network architecture is depicted in Fig.~\ref{fig:particlenetlight}. The subjets' input order and labeling are consistent with the DNN case. Both networks are optimized using the Adam optimizer.
%

\medskip
\noindent {\bf IV.~Results}

\begin{figure}[t!]
    \centering
    \includegraphics[width=1\linewidth,clip]{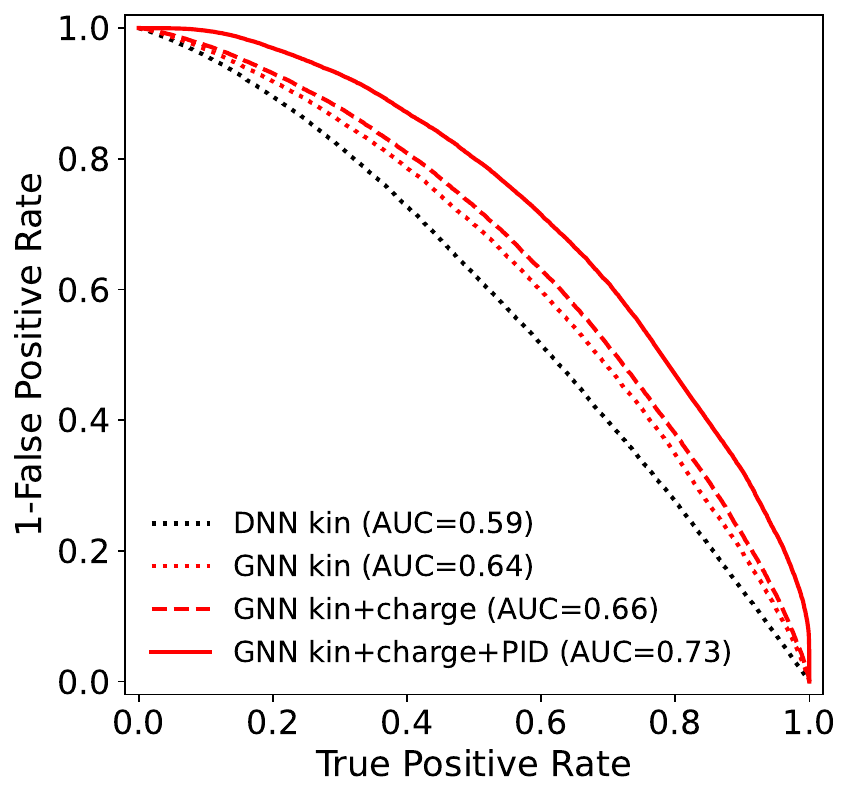}
    \caption{\label{fig.roc}Performance for tagging a down-type subjet from hadronic top quark decay with unpolarized test samples.}
\end{figure}

\noindent 
We trained each machine learning model using unpolarized data and then tested the trained models on polarized samples as well as unpolarized samples. The machine learning output scores of each model were interpreted as the probability of the soft $p(d\to q_\text{soft}|c_W,\{{\cal O}\})$ or hard subjet $p(d\to q_\text{hard}|c_W,\{{\cal O}\})$ being the down-type jet. In Fig.~\ref{fig.roc},  we present the resulting receiver operating characteristic (ROC) curves. We observe that the GNN architecture, which further explores the jet substructure and additional particle features presented in Table~\ref{table:variables}, displays significant improvements compared to the DNN, which encodes only the kinematics for the three subjets. The performance of the down-type jet classifier is summarized with the area under the ROC curve (AUC). Remarkably, the AUC value is boosted from approximately $0.59$ to $0.73$ when moving from the DNN to the GNN analysis. By examining the intermediate curves and corresponding AUCs, we can identify that the major sources of improvement stem from jet substructure kinematics and PID information.

The results for the down-type jet discrimination can be translated in terms of the spin analyzing power, using the corresponding vector length $\beta_{\rm opt}=\langle |\vec q_\text{opt}|\rangle$, defined in Eq.~\eqref{eq:opt}, as predicted by the network. Since the top quark's spin is either aligned or anti-aligned with its propagation direction in our test samples, we calculated the corresponding $\beta_{\rm opt}^{t_{L,R}}$ values after the matching process for both left and right-handed top quarks, as presented in Table~\ref{table:betas}. We observe that the GNN leads to a significant improvement in the spin analyzing power compared to the DNN baseline, which accounts only for kinematics. 

\begin{figure}[t!]
    \centering
    \includegraphics[width=1\linewidth,clip]{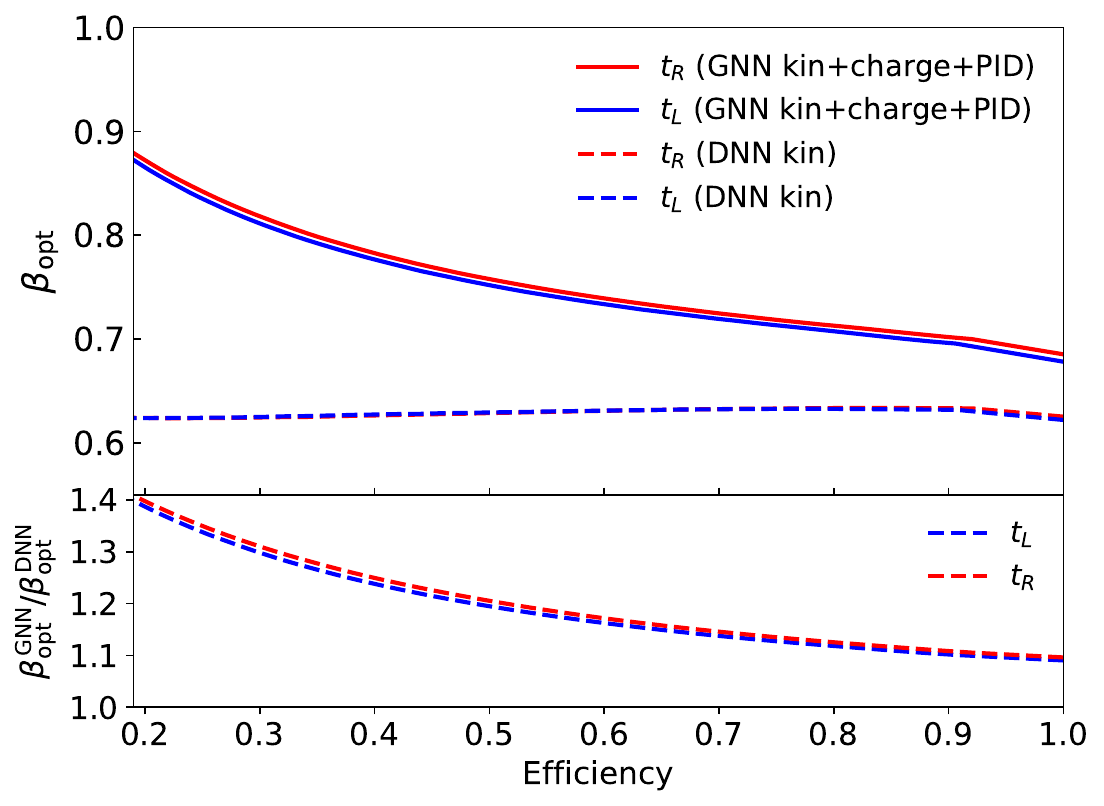}
    \caption{Spin analyzing power $\beta_{\rm opt}= \langle |\vec q_\text{opt}|\rangle$ for the left (blue line) and right-handed (red line) polarized top quarks as a function of the efficiency cuts for the GNN (solid) and DNN (dashed) output scores. We also present the ratio between the GNN and DNN results in the bottom panel. 
    \label{fig:beta_cuts}
    }
\end{figure}

These improvements can be further enhanced by increasing the purity of our samples, imposing selections on the neural network output scores. This is performed by setting two thresholds on the output scores and keeping only the events that have a score above the upper threshold or below the lower threshold. As shown in Fig.~\ref{fig:beta_cuts} and detailed in Table~\ref{table:betas}, imposing an efficiency of 50\% (20\%) can increase the spin analyzing power of the new artificial direction to $\beta^{t_{L,R}}_{\rm opt}\simeq 0.75 ~(0.86)$. This result aligns with our initial expectation that $\sigma\sim 1$ in Eq.~\eqref{eq:sigma}.

\begin{table}[!tb]
\renewcommand\arraystretch{1.3}
  \begin{tabular}{lcc}
    \hline
      & {$\beta^{t_L}_{\rm opt}$} 
      & {$\beta_{\rm opt}^{t_R}$} 
      \\
      \hline
  DNN$_\text{Eff=100\%}$  & ~~0.622~~ & ~~0.625~~ \\
  \hline
  GNN$_\text{Eff=100\%}$  & 0.678  & 0.685 \\
  GNN$_\text{Eff=50\%}$   & 0.751  & 0.758 \\
  GNN$_\text{Eff=20\%}$   & 0.863  & 0.869 \\ 
\hline
  \end{tabular}
   \caption{Spin analyzing power for different methods and efficiencies.}
\label{table:betas}
\end{table}

\medskip
\noindent {\bf V.~Conclusion}

The study of top quark polarization provides a unique and powerful tool for precision physics and new physics searches beyond the Standard Model. While the leptonic top quark decay offers a clean proxy for top quark polarization, it is statistically limited. In contrast, the hadronic top quark decays, which dominate the branching ratios, present a challenging yet potentially rich target for polarization studies.

In this work, we explored the use of jet flavor discrimination observables to enhance the sensitivity to top quark polarization in hadronic decays. We developed a down vs. up quark tagger to identify the down-type jet from hadronic top quark decays, using Graph Neural Networks. The down-type jet is then used as a proxy for top quark spin.\footnote{Since half of the hadronic top decays to charm quarks, we can further improve the up-type vs.~down-type jet discrimination  by incorporating trajectory information from the tracking system in the GNN analysis. We will leave these charm tagging improvements for future studies.}

The presented method shows that the spin analyzing power for hadronic top quarks can be improved 
by approximately 20\% (40\%) compared to the kinematic approach with an efficiency of 0.5 (0.2).
These developments not only have the potential to boost the top quark precision physics studies, but also augment the potential of top quark phenomenology for exploring physics beyond the Standard Model.

 This study opens up various possible research opportunities in connection to light-flavor jet tagging, measurement of top quark properties, spin correlations, and the exploration of entanglement and  Bell's inequalities with top quark pairs.

\medskip
\noindent {\bf Acknowledgments.}
The authors thank the organizers for the Mitchell Conference on Collider, Dark Matter, and Neutrino Physics held at Texas A\&M University, where the inspiration for this project originated.
D.G. thanks the group at the IPPP-Durham University for hosting him during the final stages of this project.
DG and AN thank the U.S.~Department of Energy for the financial support, under grant number DE-SC 0016013. 
A.L. was supported in part by the UC Southern California Hub, with funding from the UC National Laboratories division of the University of California Office of the President.
K.K. is supported by US DOE DE-SC0024407 and  Z.D. is supported in part by US DOE DE-SC0024673 and by College of Liberal Arts and Sciences Research Fund at the University of Kansas.
K.K. would like to thank the Aspen Center for Physics and the organizers of Summer 2024 workshop, ``Fundamental Physics in the Era of Big Data and Machine Learning'' (supported by National Science Foundation grant PHY-2210452) for hospitality during the completion of this manuscript.

\bibliography{draft}

\end{document}